
\input harvmac
\Title{CERN-TH 6810/93}{Large N Limit in the Quantum Hall Effect}
\centerline{ Andrea CAPPELLI\footnote{*}{\rm e-mail: cappelli@fi.infn.it,
39331::cappelli.}}
\centerline{\it I.N.F.N.}
\centerline{\it Largo E. Fermi 2, I-50125 Firenze,Italy}
\bigskip
\centerline{Carlo A. TRUGENBERGER and Guillermo R. ZEMBA}
\centerline{\it Theory Division, C.E.R.N.\footnote{**}{\rm
      e-mails: cat@vxcern.cern.ch and zemba@vxcern.cern.ch.}}
\centerline{\it 1211 Geneva 23, Switzerland}
\centerline{\bf ABSTRACT}
The Laughlin states for $N$ interacting electrons at the plateaus of the
fractional Hall effect are studied in the thermodynamic limit of large
$N$.
It is shown that this limit leads to the semiclassical regime for
these
states, thereby relating their stability to their semiclassical nature.
The equivalent problem of two-dimensional plasmas is solved analytically,
to leading order for $N\to\infty$, by the saddle-point approximation
- a two-dimensional extension of the method
used in random matrix models of quantum gravity and gauge theories.
To leading order,
the Laughlin states describe classical droplets
of fluids with uniform density and sharp boundaries,
as expected from the Laughlin ``plasma analogy''.
In this limit, the dynamical $W_\infty$-symmetry
of the quantum Hall states expresses the kinematics
of the area-preserving deformations of
incompressible liquid droplets.

\vfill
\noindent CERN-TH 6810/93

\Date{February 1993}

{\bf Introduction}

The current theory for the explanation of the plateaus
in the fractional quantum Hall effect
\ref\qhe{For a review see, {\it e.g.},
R. E. Prange and S. M. Girvin eds., {\it ``The Quantum
Hall effect''}, Springer, New York, (1990).}
is based on the seminal work of Laughlin
\ref\laugha{R. B. Laughlin, {\it ``Elementary Theory: the
Incompressible Quantum Fluid''}, in \qhe .}.
The main idea is the existence of {\it incompressible quantum
fluids} at specific rational values of the electron density.
These are very stable, macroscopical quantum states with
{\it uniform} density ${\rho({\bf x})} = \nu\ eB /hc={\rm const.}$,
$\nu=1/m, m=1,3,5,\dots$, which possess
an {\it energy gap} (here $B$ is the external magnetic field).
Incompressibility accounts for the lack of low-lying conduction
modes, which causes the longitudinal conductivity $\sigma_{xx}$
to vanish, while the overall rigid motion of the
{\it uniform droplet} of fluid gives the rational values of the Hall
conductivity $\sigma_{xy}= \nu e^{2} / h$.

In Laughlin's theory, the incompressible quantum fluid configurations
of $N$ interacting electrons are described by the wave functions
\eqn\lauwf{\psi_{m}(z_1,\dots,z_N)= \ {\cal C} \prod_{i<j=1}^{N}
{(z_i-z_j)^{m}}\
{\rm e}^{\displaystyle -{1\over 2\ell^2}\sum_{i=1}^N |z_i|^2} ,
\qquad  m=1,3,5,\dots, }
where $\ell=\sqrt{2 \hbar c / e B}$ is the magnetic length,
$\ {\cal C}=\ell^{-N-mN(N-1)/2}\ $,
and $\{ z_i ,\bar{z}_j\}$ are the coordinates of the particles in
complex notation.
The many-body properties of these states are described by expectation
values of the form
\eqn\observ{\langle{\cal O}\rangle \ = \ {1\over Z_m}\
\int\prod_{i=1}^N {d^2 z_i\over \pi\ell^2}\ \ \
{\cal O}[z_i]\ \vert\psi_m\vert^2 \ ,}
where
\eqn\zetam{Z_m\ = \ \int\prod_{i=1}^N {d^2 z_i\over \pi\ell^2}\
\exp\left\{ -\sum_{i=1}^N {\vert z_i\vert^2 \over\ell^2}\ +
m \sum_{i<j=1}^N \log {\vert z_i - z_j \vert^2 \over\ell^2} \right\}\ .}
These are interpreted as averages in the reduced statistical
problem of a {\it two-dimensional one-component plasma}
\ref\ocp{M. Baus and J.-P. Hansen, {\it Phys. Rep.} {\bf 59} (1980) 2.},
characterized by the effective temperature $T\propto 1/m$.
This {\it ``plasma analogy'' } has been an important source of physical
insight.
For example, the property $\rho({\rm x})={\rm const.}$ of the electron
ground states has been deduced \laugha\ from
the fact that the plasma is a {\it liquid } at high
effective temperatures, $m\ll 70$
\ref\caillol{J. M. Caillol, D. Levesque, J. J. Weis and
J. P. Hansen, {\it Jour. Stat. Phys.} {\bf 28} (1982) 325.}.
Debye screening, translational invariance of the liquid and numerical
results on correlation functions have been used to derive
the ground state energy
\ref\laughc{R. B. Laughlin, {\it Phys. Rev. Lett.} {\bf 50} (1983) 1395.}
, and the properties of quasi-particle excitations (like their fractional
statistics
\ref\laughb{For a review see, {\it e.g.}, R. B.
Laughlin, {\it ``Fractional Statistics in the Quantum Hall effect''},
in  F. Wilczek ed., {\it ``Fractional Statistics and Anyon
Superconductivity''}, World Scientific, Singapore, (1990).}
and their energy gaps \laugha).

A challenging problem is to devise a reliable analytic method for
computing the plasma partition function \zetam.
Similar expressions for {\it one-dimensional} plasmas are found
in the theory of random matrices
\ref\randm{For a review see, {\it e.g.}, L. M. Metha, {\it
``Random Matrices''}, Academic Press, New York, (1967).}.
Actually, Laughlin's plasma \zetam\ has been recently reformulated
as a {\it two-dimensional} matrix model
\ref\calla{D. J. Callaway, {\it Phys. Rev.} {\bf B43} (1991) 8641.}.
Given that matrix models can be solved with the $1/N$-expansion technique
\ref\largmm{E. Br\'ezin, C. Itzykson, G. Parisi and J. B. Zuber,
{\it Commun. Math. Phys.} {\bf 59} (1978) 35.},
we are lead to conclude that the same technique might be successfully
applied to Laughlin's theory.
It is the purpose of this letter to set the frame for such a large $N$
expansion in Laughlin's theory.

We derive the leading term by the {\it saddle-point approximation}
of the plasma partition function \zetam.
This correctly describes the {\it semiclassical} incompressible fluid
state of the Laughlin plasma analogy.
Our main physical point is to show explicitly that,
for the Laughlin states,
{\it the thermodynamic limit $N\to\infty$ implies the semiclassical
limit $\hbar\to 0$, and viceversa}.
While this equivalence between limits
is well-known in gauge theories and matrix models,
it acquires a different physical status in our problem, because
$N$ is a not a tunable parameter but, rather, it is naturally forced
to take large values.
Therefore, the stability of the Laughlin fluid
follows from its prominent semiclassical nature\footnote{*}{
A similar conclusion was reached in
\ref\frad{A. Lopez and E. Fradkin, {\it Phys. Rev. Lett.} {\bf 69}
(1992) 2126.}\
by a functional approach.}.

In the second part of this letter, we discuss the $W_\infty$-symmetry
underlying both Laughlin's incompressible fluids and the $c=1$ matrix
models. This symmetry has been explicitly shown to account for the
incompressibility of the $\nu=1$ Hall ground state
\ref\ctz{ A. Cappelli, C. A. Trugenberger and G. R. Zemba,{\it
``Infinite Symmetry in the Quantum Hall Effect''}, preprint
CERN-TH 6516/92, to appear in {\it Nucl. Phys.} {\bf B}.}
(similar arguments apply also to the $\nu=1/m$ Laughlin states \ctz).
Here we show that, in the large $N$ limit, the $W_\infty$
transformations reduce to the classical deformations of liquid
droplets which preserve the area.
These deformations, called {\it area-preserving diffeomorphisms},
satisfy the $w_\infty$-algebra
\ref\wref{I. Bakas, {\it Phys. Lett.} {\bf B 228} (1989) 57; C. N.
Pope, X. Shen and L. J. Romans, {\it Nucl. Phys.} {\bf B 339} (1990)
191; for a review see, {\it e.g.}: X. Shen, {\it ``W-Infinity and String
Theory''}, preprint CERN-TH 6404/92.}\footnote{**}{See also
\ref\sha{G. A. Goldin, R. Menikoff and D. H. Sharp, {\it Phys. Rev.
Lett.} {\bf 58} (1987) 2162, {\it ibid.} {\bf 67} (1991) 3499.} for a
related discussion of classical incompressible fluids.}.
This classical droplet picture has been already developed
for the $c=1$ matrix models of string theory in Refs.
\ref\wadi{A. Dhar, G. Mandal and S. R. Wadia, {\it ``Classical Fermi
fluid and geometrical action for c=1''}, IASSNS-HEP-91/89 preprint.}
\ref\sakit{S. Iso, D. Karabali and B. Sakita,
{\it Nucl. Phys.} {\bf B 388} (1992) 700, {\it Phys. Lett. } {\bf B 296}
(1992) 143.},
where the relation to the Landau level problem has been also recognized.
By identifying the $\hbar\to 0$ and the $N\to\infty$ limits,
we show that this picture nicely fits into Laughlin's plasma analogy.

\bigskip
\bigskip
{\bf Saddle-point approximation}

Let us start by recalling some known facts about the plasma partition
function $Z_m$ in eq.\zetam. For $m=1$, it can be computed exactly by
using the orthogonality of the first Landau level wave functions,
\eqn\landwf{\varphi_{k}(z,{\bar z})\ =\ {1 \over {\ell\sqrt{\pi k !}}}\
\left({z \over \ell}\right)^{k}\ {\rm e}^{-|z|^2/ 2{\ell}^2 } \ ,}
where $k$ is the angular momentum eigenvalue.
The result is
\eqn\logzone{\eqalign{\log Z_1\ &=\ \log \left( N! \prod_{n=0}^{N-1}
n!\right) \cr
&=\ {N^2 \over 2}\left(\log N -{3\over 2}\right)\ +\ \log N!\ +\
{N\over 2}\log 2\pi\ -\ {1\over 12}\log N\ +\ O(1)\ .\cr}}
We are interested in the observable one-particle density,
\eqn\rhoone{\rho_1 ({\bf x})\ \equiv\ \langle\Omega_1\vert\Psi^{\dag}
({\bf x})\Psi({\bf x})\vert\Omega_1\rangle\ ,}
where $\vert\Omega_1\rangle$ is the $\nu=1$ ground state and
$\Psi$ the field operator in fermionic Fock space,
\eqn\psif{\Psi (z,{\bar z})\ = \sum_{k=0}^{\infty}
F_{k} \ \varphi_{k} (z,{\bar z}),\qquad
\qquad \{ F_{k}, F^{\dag}_{l} \}= \delta_{k,l}\ .}
The density is easily computed:
\eqn\exprho{\rho_{1}(z,{\bar z})\ =\
{1\over \ell^2 \pi}\ {\rm e}^{-r^2/\ell^2 }\ \sum_{k=0}^{N-1}
{1\over k!} \left(  {r\over\ell} \right)^{2k}\ ,\qquad r\equiv|z|
\ ,} and is plotted in Fig. 1 for $N=50$. It is constant
for $r \ll  \ell\sqrt{N}$,
and drops rapidly to zero around $r \simeq \ell\sqrt N$.
This is the density profile of a {\it quantum} droplet of incompressible
fluid. The fluid character is reflected by the uniform value in the
interior, and incompressibility follows from the gap for density
fluctuations (the cyclotron energy).
Quantum behaviour is apparent from the {\it smooth} boundary, where the
occupation probability is neither zero nor one.
In contrast, the density of a classical droplet of liquid has a
{\it sharp} boundary.

Actually, {\it the large $N$ limit of the quantum density \exprho\ is a
classical density}.
Let us introduce the rescaled coordinate
\eqn\scalv{w \ =\ {z \over\sqrt{N}}\ ,}
and the rescaled density $\ \rho_1(w)\equiv \rho_1(|z|=\sqrt{N}|w|)\ $,
satisfying $\ \int d^2w\ \rho_1(w)=1\ $.
The rescaled density possesses a finite large $N$ limit,
\eqn\limn{\lim_{N\to\infty}\
\rho_1 \left(\vert z\vert = \sqrt{N} \vert w\vert \right)
\ = \ {1\over \pi\ell^2} \ \Theta
\left( 1 -{\vert w\vert^2\over\ell^2} \right)\ ,}
where $\Theta$ is the step-function. The sharpness of the boundary
is a first indication of the classical nature of the $N\to\infty$
limit.
In eq. \limn , $\ell^2$ has to be understood as the classical
parameter setting the scale for the electron density through
\eqn\andr{{N\over A}= \nu {B\over \Phi_0} =\nu {B\over (hc/e)} =
\nu {1\over \pi\ell^2} \ ,}
for uniform filling $\nu$ and given area $A$ of the sample.

Equation \andr\ implies that
the naive semiclassical limit $\hbar\to 0$ with all the other parameters
fixed cannot be taken in our problem.
For a given external magnetic field $B$ and type of fluid characterized
by $\nu$, the limit $\hbar \to 0$ enforces
$N\to \infty$ if the system is to have a finite macroscopical
area $A$.
This is another general argument for the advertised equivalence of
the $N\to\infty$ and $\hbar\to 0$ limits.

No analytic expression is known for the densities $\rho_m$,
$m=3,5,\dots$, corresponding to the Laughlin states \lauwf.
 Numerical studies
\ref\morf{R. Morf and B. I. Halperin, {\it Phys. Rev.} {\bf B33}
(1986) 2221.}
\ref\ferra{N. Datta and R. Ferrari, Max-Planck Institute preprint
MPI-Ph/92-16, March 1992.}
for large number of particles (up to $N=200$), show a
constant density $\ \rho_{m} = 1/m\pi\ell^2\ $ for
$\ r \ll  \ell\sqrt{mN}\ $, followed by an upward bump near the
boundary region, where $\rho_{m}$ drops rapidly to zero. The fact
that this bump does not seem to decrease rapidly when the particle
number $N$  is increased up to $N=200$ has led the authors
of \ferra\ to conclude that Laughlin's wave function does not describe
a uniform quantum fluid.
Actually, we are now going to show that this is not the case.
Indeed, the limiting large $N$ form of the densities $\ \rho_m\ $
will be shown analytically to be:
\eqn\rhorot{\lim_{N\to\infty}\rho_{m} \left(|z|=\sqrt{N} |w|\right)\ =\
{1\over m \pi\ell^2}\ \Theta\left( m - {|w|^{2}\over\ell^2} \right)\ .}
These describe again classical droplets of incompressible fluid;
therefore, the bump seen in ref.\ferra\ has to disappear
for $N\to\infty$.

This result can be obtained by extending the saddle-point technique
of Brezin, Itzykson, Parisi and Zuber \largmm\ to $Z_m$ in eq.\zetam
\footnote{*}{See also
\ref\koga{I. Kogan, A. M. Perelomov and G. W. Semenoff,
{\it Phys. Rev.} {\bf B45} (1992) 12084.}
for a similar approach.}.
We first rewrite:
\eqn\zetap{\eqalign{ Z_m\ &=\ \int\prod_{i=1}^N {d^2 z_i\over \pi\ell^2}\
\exp\left(-N H_m [w] \right)\ ,\cr
H_m[w]\ &=\  \sum_{i=1}^N {\vert w_i\vert^2 \over\ell^2}\  -\ {m\over N}\
\sum_{i<j=1}^N \log {\vert w_i - w_j \vert^2 \over\ell^2}\  .\cr} }
For large $N$, the particles are driven
into a saddle point configuration $\{ w_i=w_i^0 \}$, determined by the
equation
\eqn\sadeq{{\bar w}_{i}\ =\ {m\over N}\ \ell^2\sum_{j,j \neq i}
{1\over{w_i - w_j}}\ .}
By considering a lattice decomposition of the plane,
we can replace the sum over particles with the sum over cells times
the characteristic function for cell occupation.
For $N\to\infty$, the latter becomes a continuous distribution for the
rescaled variable $w$, which equals the electron density
$\ \rho_m (w)\ $ to leading order.
Therefore, we perform the replacement
\eqn\cont{\sum_i
\to \int d^{2} z\ \rho_{m}(z)\ =\ N \ \int d^{2} w\ \rho_{m} (w)\ .}
The saddle-point equation \sadeq\  becomes the integral
equation\footnote{**}
{The point $w=w'$ excluded in the sum \sadeq\ causes no harm in the
following integral for continuous $\rho$ functions.}
\eqn\specont{{\bar w}\ =\ m \ \ell^2
\int d^{2} w'\ {\rho_{m}(w')\over{w-w'}}\ ,}
whose solutions are subjected to the normalization
\eqn\normrho{\int d^{2} w\ \rho_{m}(w)\ =\ 1\ .}

The double integration makes the integral equation \specont\ more
involved than the corresponding one for
one-dimensional matrix models \largmm\ -
no general solution is known to us.
However, it is easy to check that the rotational invariant density
\rhorot\ is a solution\footnote{***}{
To see this, it is easier to perform first
 the angular integration using the theorem of residues.}.
Furthermore, the following argument shows that this solution is stable,
{\it i.e.}, it is a local minimum of $H_m$ in eq.\zetap.
By scaling the variable $w=\sqrt{m} q$, one obtains the identity
\eqn\zetaid{Z_m \ = \ \exp\left( {N(N-1)\over 2} m\log m \right)\
\left(Nm\right)^N\ \int\prod_i {d^2 q_i\over\pi\ell^2}\ \
e^{-mNH_1[q]}\ ,}
which can be approximated semiclassically to second order, yielding
\eqn\quadra{ Z_m \sim \ {\rm e}^{-NH_m[w^0]  }
\ N^N\int\prod_i {d^2 \ \delta w_i\over\pi\ell^2}\ \
\exp \left({ -{N\over 2}\sum_{i,j}\  \delta v_i
{\partial^{2}H_{1}[w^0] \over\partial v_{i} \partial v_{j} }
\delta v_j}\right) \ ,}
where $\ v_i=(w_i,\bar w_i)\ $.
This equation shows that quadratic fluctuations are
{\it independent of m}.
Moreover, for $m=1$ the saddle-point solution is clearly
quadratically stable,
since it agrees with the exact solution \exprho\limn.
Therefore, the solution \rhorot\ is stable for any $m$.

The value of the Hamiltonian at these saddle points can also be easily
computed in the continuum approximation and it reads:
\eqn\logzm{\log Z_m ={N^2m\over 2}\left( \log Nm -{3\over 2}\right)
\ +\ \log N! \ +\ \dots\ ,}
which also matches smoothly the exact value for $m=1$ in eq.\logzone\
to leading order.

The occurrence of other stable solutions of lower ``energy'' than \logzm\
for $m=3,5,\dots ,$
is not excluded by our analysis, but it is very unlikely for small values
of $m$.
Numerical simulations of the two-dimensional plasma \caillol\ indicate
a phase transition to a Wigner crystal only at a large value
$m_{crit}\sim 70$.
It would be very interesting to find the semiclassical
solution with broken rotational invariance corresponding to the Wigner
crystal.
Let us also remark that the one-dimensional plasma
of matrix models has only one phase \randm.

In conclusion, for $N\to\infty$ we confirm that the Laughlin wave
functions \lauwf\ describe droplets of uniform fluid with densities
$\ \rho_m=1/m\pi\ell^2\ $ and sharp boundaries.
The interesting structure at the boundary of the droplets found by the
numerical calculations \ferra\ might be related to {\it edge excitations}
\ref\wenrev{B. I. Halperin, {\it Phys. Rev.} {\bf B25} (1982) 2185;
X. G. Wen, {\it ``Gapless Boundary Excitations in the Quantum
Hall States and the Chiral Spin States''}, preprint NSF-ITP-89-157,
{\it Phys. Rev. Lett.} {\bf 64} (1990) 2206;
M. Stone, {\it Ann. Phys.} (NY) {\bf 207} (1991) 38;
J. Fr\"ohlich and T. Kerler, {\it Nucl. Phys} {\bf B 354} (1991) 369;
for a review see: X. G. Wen, {\it Int. Jour. Mod. Phys. }
{\bf B6} (1992) 1711.},
which are subleading $O(1/N)$ boundary effects
\ref\cdtz{A. Cappelli, G. V. Dunne, C. A. Trugenberger and G. R. Zemba,
{\it ``Conformal Symmetry and Universal Properties of Quantum Hall
States''}, preprint CERN-TH 6702/92, {\it Nucl. Phys.} {\bf B}, in
press.}.

\bigskip
\bigskip
{\bf Observables to leading order}

Some physical information on quasi-particle excitations can be obtained
by evaluating observables \observ\ within the saddle-point approximation.
More precisely, all Laughlin's results based on the plasma analogy
can be rephrased in this approximation.
Following the review article \laughb\ we can, {\it e.g.}, verify the
normalization of the wave function for one quasi-hole at the point $z$,
\eqn\qhwf{\psi_{QH}(z;z_1,\dots,z_N)= \ e^{-|z|^2 /2m\ell^2}\
\prod_{i=1}^N (z- z_i)\ \psi_{m} (z_1,\dots,z_N)\ \equiv
S[z,z_i]\ \psi_m (z_1,\dots,z_N)\  ,}
with $\psi_{m}$ the Laughlin wave function \lauwf .
Let us compute $\Vert \psi_{QH}\Vert^2\ $, {\it i.e.} eq.\observ\ for the
operator ${\cal O}=S^{\dag} S$.
Repeating the previous steps for the modified plasma with one
added charge,
we find that the saddle-point equation \specont\ is not modified
to leading
order, and that the saddle-point value of this observable is
indeed unity.
Actually, the saddle-point solution \rhorot\ is not modified by the
inclusion of any finite number of charges.

The wave function of two quasi-holes \laughb\ can be treated similarly.
One finds that the correlation among them, showing their fractional
statistics, is a subleading effect.
This agrees with the result of the theory of edge excitations \wenrev,
in which it has been shown that quasi-particle correlations are of order
$O(1/N)$ \cdtz.
As expected, quasi-particles, which are quantum effects, are not seen to
leading large $N$ order.

\bigskip
\bigskip
{\bf Droplet picture and ${\bf w}_\infty$ symmetry}

So far, we have been considering the large $N$ limit of the density
from a computational point of view. Now, we would like to discuss
the {\it geometrical} interpretation of this limit.

Before doing that, let us recall the $W_\infty$ dynamical {\it quantum}
symmetry of the $\nu=1$ ground state recently found in ref.\ctz.
There, we constructed operators ${\cal L}_{n,m}$,
living in the first Landau level,
\eqn\defl{{{\cal{L}}_{n,m}} \equiv \sum_{i=1}^N\
(b_i^{\dag})^{n+1}\ b_i^{m+1},
\qquad n,m \geq -1, }
where $b_i,b_i^{\dag}$ are the harmonic oscillators for angular momentum
$J$ excitations,
\eqn\bbdag{b_i ={\bar z_i \over{2\ell}}+{\ell}\partial_i  \ ,
\qquad b_i^{\dag } ={z_i \over{2\ell}}-{\ell}{\bar\partial_i} \ ,\qquad
[b_i, b_j^{\dag }]=\delta_{ij}\ , \qquad
J\ =\ \sum_i\ b^{\dag}_i b_i \ .}
The ${\cal L}_{n,m}$ satisfy the $W_\infty$ algebra\footnote{*}{
More precisely, this is a $W_{1+\infty}$ algebra.}
\eqn\em{ \eqalign{[{{\cal{L}}_{n,m}},{{\cal{L}}_{k,l}}] &=\ \left(
\sum_{s=0} ^{Min(m,k)} { (m+1)! (k+1)! \over{(m-s)! (k-s)! (s+1)!}} \
{\cal{L}}_{n+k-s,\ m+l-s} \right)\ -\ \left(
m \leftrightarrow l ,n \leftrightarrow k\right)\ \cr
\ &= \left( (m+1)(k+1)-(n+1)(l+1) \right ) {{\cal{L}}_{n+k,m+l}} \ +\
(\dots)\ {\cal L}_{n+k-1,m+l-1} +\dots . \cr} }
In particular, the angular momentum is $J={\cal L}_{00}$.
The subalgebra
\eqn\lzero{[{\cal L}_{00},{\cal L}_{n,m}] = (n-m){\cal L}_{n,m} \ ,}
shows that the ${\cal L}_{n,m}$ are raising $(n>m)$ and
lowering $(n<m)$ operators for angular momentum.

The quantum symmetry of the $\nu=1$ ground state is encoded in the
invariance of the ground state under an infinite (for $N\to\infty$) set
of these transformations, {\it i.e.}, the highest-weight conditions \ctz
\eqn\hwc{{\cal L}_{n,m}\ \psi_1(z_1,\dots,z_N)\ =\ 0, \qquad n<m\ .}
These are interpreted as the algebraic conditions of
incompressibility, since eq.\hwc\ means that all transitions
lowering the angular momentum of the ground state,
{\it i.e., compressions}, are impossible.

Next, we discuss the corresponding classical picture.
As shown before, when $N\to\infty$, the quantum density
of the Laughlin states at $\nu=1/m$ reduces to the profile of a classical
droplet of liquid with uniform density $\ \rho_m=1/m\pi\ell^2\ $.
Consider now a deformation of this droplet.
The density cannot change locally, due to incompressibility.
Moreover, its space integral gives the particle number $N$, and is,
therefore, constant.
Thus, the {\it area} occupied by the droplet stays constant, {\it i.e.},
deformations can only produce droplets of the same area and
different shapes.
Therefore, the different configurations of the incompressible fluid
are related by {\it area-preserving diffeomorphisms},
whose generators satisfy the $w_\infty$ algebra \wref.
In the following, we give an explicit derivation of the action
of $W_\infty$ on the observable density and its classical limit
$w_\infty$, thereby confirming this picture.

To this end we use a different basis for the
generators of $W_{\infty}$. By using this new basis, we will stress
the analogy to
the parallel discussion of the $\hbar\to 0$
limit of the $c=1$ matrix model in $(1+1)$ dimensions, as formulated in
\wadi\sakit, which possesses the same $W_\infty$ symmetry.
The basic reason for this analogy is that the Hilbert spaces of the two
systems are isomorphic.
In the matrix model, one considers the Hamiltonian
$H={1\over 2} \left( p^2 - x^2 \right)$
yielding a real representation of the harmonic Fock space.
In the first Landau level, we have the holomorphic representation \bbdag,
where angular momentum plays the role of the Hamiltonian.
As is well known, in Bargmann space the two coordinates $(z, \bar z)$
become conjugate variables of a $(1+1)$-dimensional phase space
\ref\itz{For a review, see: C. Itzykson, {\it ``Interacting electrons
in a Magnetic Field''} , in {\it ``Quantum Field Theory and
Quantum Statistics,
Essays in Honor of 60th Birthday of E. S. Fradkin''},
A. Hilger, Bristol, (1986).}.

In analogy with \wadi , we consider the Wigner
phase-space distribution:
\eqn\wigdf{W(k,{\bar k})\equiv\int d^2 z\ \Psi^{\dag}(z)\
{\rm e}^{{i\over 2}\ell \left( \bar k b^{\dag} + kb\right)}\ \Psi (z)\
=\ e^{{\ell^2}k{\bar k}/ 8 }\ \int d^{2} z\ \Psi^{\dag}(z)\
{\rm e}^{\scriptstyle +{{i\over 2}({\bar k}z+k{\bar z})}} \ \Psi (z)\ .}
This is the generating function for the operators\footnote{*}{
In equation \wigdf, $\Psi$ is the field operator \psif;
the ${\cal L}_{n,m}$ appearing hereafter are, therefore, expressed in the
second quantized formulation \ctz.}
$\ {\cal L}_{n,m}$ in eq.\defl,
\eqn\wilnm{ W(k,{\bar k})\ =\
\sum_{n,m=0}^{\infty}
{1\over{n!m!}} \left( {i\ell {\bar k} \over{2}} \right )
^{n} \left( {i\ell  k \over{2}} \right )^{m} :{\cal{L}}_{n-1,m-1}:\ ,}
where $:\ :$ denotes Weyl normal ordering \wadi , {\it e.g.},
$:(b^{\dag})^{2} b:\equiv {1\over 3}\left((b^{\dag})^{2} b + b^{\dag} b
b^{\dag} + b (b^{\dag})^{2} \right)$.
This shows that the
generating functions $W(k,{\bar k})$ form a different basis for the
$W_{\infty}$ generators. In terms of these new generators,
the $W_\infty$ algebra \em\ acquires the compact form
\eqn\comwig{[W(k,{\bar k}),W(p,{\bar p})]\ =\
2 \sinh \left( {{\ell^2}\over 8} (p{\bar k}-{\bar p}k) \right)
\ W(p+k,{\bar p}+ {\bar k})\ .}
This result is analogous to the $(1+1)$-dimensional result of \wadi,
due to the previously noticed mapping between Hilbert spaces.

The Fourier transform of $W(k,\bar k)$ is the one-particle density
{\it operator} of eq.\rhoone,
\eqn\exprho{\rho(z,{\bar z})\ =\ \int {{d^{2} k} \over (2\pi)^2} \
{\rm e}^{\scriptstyle {-{i\over 2}(k{\bar z}+{\bar k}z) }}\
W(k,{\bar k})\ {\rm e}^{\scriptstyle -{{\ell^2}k{\bar k}/8}}\ =\
\Psi^{\dag}(z) \Psi (z) \ ,}
apart from a normal-ordering factor.
We recall that, originally
\ref\wign{E. P. Wigner, {\it Phys. Rev.} {\bf 40} (1932) 749.},
Wigner distributions were introduced as the quantum
analogs of phase-space distributions of classical
statistical mechanics. This is because quantum expectation values
can be written as classical averages in phase space with these
distributions.
In our Landau level problem, the phase space is given by $(z,\bar z)$;
thus a phase space distribution is actually a two-dimensional space
distribution, as in \exprho.

The action of a $W_\infty$ transformation generated by $W(k,\bar k)$,
with infinitesimal parameter $\epsilon (k,\bar k)$, on the ground state
density \rhoone\ is given by
\eqn\defrho{\delta_{\epsilon}\ \rho_1 (z,{\bar z})\ =\
\int d^{2}k\ \epsilon (k,{\bar k})\ \ \delta_{k,{\bar k}}\
\rho_1 (z,{\bar z})\ ,}
where
\eqn\wact{\eqalign{\delta_{k,{\bar k}} \rho_1 (z,{\bar z})
&\equiv\ i\langle\Omega_1|[\rho_1 , W(k,{\bar k})]|\Omega_1 \rangle\ \cr
&=\ i{\rm e}^{\displaystyle \ {{\ell^2}k{\bar k}\over 8}} \left(
{\rm e}^{\displaystyle \ {i\over 2}{\ell^2} k {\partial\over\partial z} }
\ -\ {\rm e}^{\displaystyle \ {i\over 2}{\ell^2}
        {\bar k} {\partial\over\partial {\bar z}} } \right )
{\rm e}^{\displaystyle \ {i\over 2} (k{\bar z}+{\bar k}z) }
\rho_1 (z,{\bar z})\ .\cr}}
This expression becomes more transparent in the large $N$ limit,
which is achieved, as before, by introducing the rescaled coordinate
$w=z/\sqrt{N}$, and correspondingly, the rescaled
momentum $\kappa =k\sqrt{N}$.
After rescaling, eq. \wact\ reads,
\eqn\wscal{\eqalign{\delta_{\kappa,{\bar\kappa}}\ \rho_1 (w,{\bar w})
&=\ i\langle\Omega_1|
   \left[\rho_1 , W\left( {\kappa\over\sqrt{N}},{\bar\kappa\over\sqrt{N}}
     \right)\right] |\Omega_1 \rangle\ \cr
&=\ i{\rm e}^{\displaystyle \ {{\ell^2}\over N} {\kappa{\bar\kappa}\over
8}}\left(
{\rm e}^{\displaystyle \ {i{\ell^2}\over 2N}
\kappa {\partial\over\partial w} }
\ -\ {\rm e}^{\displaystyle\ {i{\ell^2}\over 2N} {\bar\kappa}
             {\partial\over\partial {\bar w}} } \right )
{\rm e}^{\displaystyle \ {i\over 2} (\kappa{\bar w}+{\bar\kappa} w) }
\rho_1 (w,{\bar w})\ ,\cr}}
where we recall that
$\ell^2 /N = 2\hbar c/eBN$.
In the large $N$ limit, eq. \wscal\ reduces to
\eqn\wlim{\delta^{(cl)}_{\kappa,{\bar\kappa}} \rho_1 (w,{\bar w})\ =\
\left\{\ \rho_1(w,\bar w)\ , \ -
{\rm e}^{{i\over 2} (\kappa{\bar w}+{\bar\kappa} w) }\right\}_{PB}\ ,}
where
$\delta^{(cl)}_{\kappa,{\bar\kappa}}\ \equiv\ {N \over{\hbar}}
\delta_{\kappa,{\bar\kappa}}$, and
$\{ f ,g \}_{PB}\equiv -{i\over{\cal B}} \left(
(\partial f/\partial w) (\partial g/\partial\bar w)
-(\partial f/\partial \bar w) (\partial g/\partial w) \right)$,
with ${\cal B} \equiv  eB/2c$,
denotes the correct Poisson bracket
with respect to the classical variables $w$ and $\bar w$.
This equation shows that, to leading large $N$ order,
$W\left(\kappa/\sqrt{N}\ ,\bar\kappa/\sqrt{N}\ \right)$
is the generating function
of canonical, and therefore area-preserving, transformations in the
two-dimensional phase space $(w,{\bar w})$.
Eq. \wlim\ makes manifest the {\it classical nature} of a
$W_{\infty}$-transformation acting on the density, in the {\it large $N$
limit}.
As we now show, in this limit the quantum $W_\infty$ algebra reduces to
the classical algebra of area-preserving diffeomorphisms $w_\infty$.
Our result matches the $\hbar\to 0$ limit of Refs. \wadi\sakit,
thereby further confirming that the $N\to\infty$ is a semiclassical
limit.

We now verify the $w_\infty$ algebra for the classical analogs of
the operators $W(k,\bar k)$ and ${\cal L}_{n,m}$ acting by Poisson
brackets on functions of the holomorphic phase space.
To this end, we perform the coordinate and momentum rescalings as before,
and we use the correspondence between operators and classical
{\it functions},
$\ N\ W\left( \kappa/\sqrt{N}\ ,\bar \kappa/\sqrt{N}\right) \to
W^{(cl)} (\kappa,\bar\kappa)\ $
and $\ [\ ,\ ] \to i{\hbar}\{\ ,\ \}_{PB}$.
The classical limit of eq.\comwig\ reads
\eqn\cwig{
\left\{ W^{(cl)}(\kappa,\bar\kappa),
W^{(cl)}(\lambda,\bar\lambda) \right\}_{PB}
= -{i\over{4{\cal B}}} \left( \lambda\bar\kappa -\bar\lambda\kappa
\right)\ W^{(cl)}(\kappa+\lambda,\bar\kappa +\bar\lambda)\ ,}
which verifies the identification
$W^{(cl)}(\kappa,\bar\kappa)=
-e^{{i\over 2}(\kappa \bar w+\bar\kappa w)}$ from eq. \wlim.
This, in turn, identifies the classical limit of the operators
${\cal L}_{n,m}$ through eq.\wilnm,
\eqn\clnm{\ell^2\left({\ell \over{\sqrt{N}}}\right)^{n+m}\ {\cal L}_{n,m}\
\to\ {\cal L}^{(cl)}_{n,m}\ = \ -w^{n+1}\bar w^{m+1} \ .}
Their classical algebra is
\eqn\clem{ \{ {{\cal{L}}^{(cl)}_{n,m}},{{\cal{L}}^{(cl)}_{k,l}}
\}_{PB}\ = - {i\over{\cal B}} \left( (m+1)(k+1)-(n+1)(l+1) \right )
{{\cal{L}}^{(cl)}_{n+k,m+l}}\ ,}
which agrees with the classical limit of eq.\em.
Equations \cwig\ and \clem\ are equivalent forms for the $w_\infty$
algebra of area-preserving diffeomorphisms.
Note that the classical functions ${\cal L}^{(cl)}_{n,m}$
and $W^{(cl)}$ and their algebras \cwig\ and \clem\
can be also derived directly from the canonical treatment
of the classical theory describing the dynamics of the first Landau
level, the ``topological quantum mechanics'' of refs.
\ref\topcs{G. V. Dunne, R. Jackiw and C. A. Trugenberger, {\it Phys.
Rev.} {\bf D 41} (1990) 661; G. V. Dunne and R. Jackiw,
MIT preprint CTP 2123 (June 1992), to appear in Umezawa volume.}
(see also \ctz).

Having identified the classical
${\cal L}^{(cl)}_{n,m}$, we can evaluate their action on the
classical density \limn\ by using eq. \wlim :
\eqn\wlima{\eqalign{\delta^{(cl)}_{n,m} \ \rho_1 (w,{\bar w})\
 &=\ \left\{\ \rho_1(w,\bar w)\ , \ w^{n+1} {\bar w}^{m+1}
\right\}_{PB}\ \cr
 &=\ i (n-m)\ {{\ell}^{n+m-2} \over{\pi {\cal B}}}\
 {\rm e}^{i(n-m)\theta}\ \delta\left(
1-{|w|^{2} \over{\ell^2}} \right)\ ,\cr}}
where $\theta$ is the angular variable on the circle delimiting the
classical droplet. These variations correspond to {\it density waves
localized on the one-dimensional sharp boundary of the classical
droplet}. The quantization of these edge waves leads to a
$(1+1)$-dimensional $c=1$ conformal field theory \wenrev \cdtz.

So far, our explicit derivation of the classical $w_\infty$ algebra
referred
only to the $\nu=1$ ground state; however, it can be easily extended to
the Laughlin states at $\nu=1/m$. Indeed, their classical densities
$\ \rho_m\ $ are related to $\ \rho_1\ $ by the simple scaling
\hbox{ $\rho_m(w)=(1/m)\rho_1(w/\sqrt{m})$}.
Therefore, we can apply the same scaling
in the classical phase space to obtain the corresponding $w_\infty$
action \wlim-\clem\ for $\nu=1/m$.
We can also argue that the dynamical quantum symmetry of the Laughlin
states must be the quantum extension $W_\infty$.
Actually, the analogs of the symmetry conditions \hwc\ were found
in Ref.\ctz. However, in going from $w_\infty$ to $W_\infty$,
one needs the explicit form of the higher order corrections
$O(1/N^k), k>0$ in the algebra $W_\infty$, as in eq.\em.
These are not unique and their correct form is not presently known
to us for $\nu=1/m$.

\bigskip
\bigskip
{\bf Acknowledgments}
\bigskip
\noindent
G. R. Z. thanks the partial support by the World Laboratory and the
hospitality of the Physics Department of the University of Firenze.
A.C. thanks the hospitality of the Theory Group at CERN where this
work done.
\listrefs

{\bf Figure caption}

{\bf Fig. 1}

The density profile in units of $1/\pi\ell^2$ for the first Landau level
filled up to $L=50$ as a function of $r/\ell$.

\end